\RequirePackage[mathlines]{lineno}
\documentclass[floatfix,twocolumn,showpacs,preprintnumbers,amsmath,amssymb,prc,superscriptaddress]{revtex4}

\usepackage{graphicx}
\usepackage{dcolumn}
\usepackage{bm}

\def\sqrtsNN{\mbox{$\sqrt{s_\mathrm{NN}}$}}

\newcommand{ \be }{\begin{equation}}
\newcommand{ \ee }{\end{equation}}
\newcommand{ \bea }{\begin{eqnarray}}
\newcommand{ \eea }{\end{eqnarray}}    {\Large {\normalsize }}
\newcommand{ \la }{\langle}
\newcommand{ \ra }{\rangle}

\usepackage{color}

\usepackage{ulem}

\begin{document}

\title{Directed Flow of Identified Particles in Au + Au Collisions at $\sqrtsNN = 200$ GeV at RHIC}

\affiliation{Argonne National Laboratory, Argonne, Illinois 60439, USA}
\affiliation{Brookhaven National Laboratory, Upton, New York 11973, USA}
\affiliation{University of California, Berkeley, California 94720, USA}
\affiliation{University of California, Davis, California 95616, USA}
\affiliation{University of California, Los Angeles, California 90095, USA}
\affiliation{Universidade Estadual de Campinas, Sao Paulo, Brazil}
\affiliation{Central China Normal University (HZNU), Wuhan 430079, China}
\affiliation{University of Illinois at Chicago, Chicago, Illinois 60607, USA}
\affiliation{Krakow Universities and Institute}
\affiliation{Creighton University, Omaha, Nebraska 68178, USA}
\affiliation{Czech Technical University in Prague, FNSPE, Prague, 115 19, Czech Republic}
\affiliation{Nuclear Physics Institute AS CR, 250 68 \v{R}e\v{z}/Prague, Czech Republic}
\affiliation{University of Frankfurt, Frankfurt, Germany}
\affiliation{Institute of Physics, Bhubaneswar 751005, India}
\affiliation{Indian Institute of Technology, Mumbai, India}
\affiliation{Indiana University, Bloomington, Indiana 47408, USA}
\affiliation{Alikhanov Institute for Theoretical and Experimental Physics, Moscow, Russia}
\affiliation{University of Jammu, Jammu 180001, India}
\affiliation{Joint Institute for Nuclear Research, Dubna, 141 980, Russia}
\affiliation{Kent State University, Kent, Ohio 44242, USA}
\affiliation{University of Kentucky, Lexington, Kentucky, 40506-0055, USA}
\affiliation{Institute of Modern Physics, Lanzhou, China}
\affiliation{Lawrence Berkeley National Laboratory, Berkeley, California 94720, USA}
\affiliation{Massachusetts Institute of Technology, Cambridge, MA 02139-4307, USA}
\affiliation{Max-Planck-Institut f\"ur Physik, Munich, Germany}
\affiliation{Michigan State University, East Lansing, Michigan 48824, USA}
\affiliation{Moscow Engineering Physics Institute, Moscow Russia}
\affiliation{Ohio State University, Columbus, Ohio 43210, USA}
\affiliation{Old Dominion University, Norfolk, VA, 23529, USA}
\affiliation{Panjab University, Chandigarh 160014, India}
\affiliation{Pennsylvania State University, University Park, Pennsylvania 16802, USA}
\affiliation{Institute of High Energy Physics, Protvino, Russia}
\affiliation{Purdue University, West Lafayette, Indiana 47907, USA}
\affiliation{Pusan National University, Pusan, Republic of Korea}
\affiliation{University of Rajasthan, Jaipur 302004, India}
\affiliation{Rice University, Houston, Texas 77251, USA}
\affiliation{Universidade de Sao Paulo, Sao Paulo, Brazil}
\affiliation{University of Science \& Technology of China, Hefei 230026, China}
\affiliation{Shandong University, Jinan, Shandong 250100, China}
\affiliation{Shanghai Institute of Applied Physics, Shanghai 201800, China}
\affiliation{SUBATECH, Nantes, France}
\affiliation{Texas A\&M University, College Station, Texas 77843, USA}
\affiliation{University of Texas, Austin, Texas 78712, USA}
\affiliation{University of Houston, Houston, TX, 77204, USA}
\affiliation{Tsinghua University, Beijing 100084, China}
\affiliation{United States Naval Academy, Annapolis, MD 21402, USA}
\affiliation{Valparaiso University, Valparaiso, Indiana 46383, USA}
\affiliation{Variable Energy Cyclotron Centre, Kolkata 700064, India}
\affiliation{Warsaw University of Technology, Warsaw, Poland}
\affiliation{University of Washington, Seattle, Washington 98195, USA}
\affiliation{Wayne State University, Detroit, Michigan 48201, USA}
\affiliation{Yale University, New Haven, Connecticut 06520, USA}
\affiliation{University of Zagreb, Zagreb, HR-10002, Croatia}

\author{L.~Adamczyk}\affiliation{Krakow Universities and Institute}
\author{G.~Agakishiev}\affiliation{Joint Institute for Nuclear Research, Dubna, 141 980, Russia}
\author{M.~M.~Aggarwal}\affiliation{Panjab University, Chandigarh 160014, India}
\author{Z.~Ahammed}\affiliation{Variable Energy Cyclotron Centre, Kolkata 700064, India}
\author{A.~V.~Alakhverdyants}\affiliation{Joint Institute for Nuclear Research, Dubna, 141 980, Russia}
\author{I.~Alekseev}\affiliation{Alikhanov Institute for Theoretical and Experimental Physics, Moscow, Russia}
\author{J.~Alford}\affiliation{Kent State University, Kent, Ohio 44242, USA}
\author{B.~D.~Anderson}\affiliation{Kent State University, Kent, Ohio 44242, USA}
\author{C.~D.~Anson}\affiliation{Ohio State University, Columbus, Ohio 43210, USA}
\author{D.~Arkhipkin}\affiliation{Brookhaven National Laboratory, Upton, New York 11973, USA}
% ...

\collaboration{STAR Collaboration}\noaffiliation

\begin{abstract}
STAR's measurements of directed flow ($v_1$) around midrapidity for
$\pi^{\pm}$, K$^{\pm}$, K$_S^0$, $p$ and $\bar{p}$ in Au + Au
collisions at $\sqrtsNN = 200$ GeV are presented. A negative
$v_1(y)$ slope is observed for most of produced particles
($\pi^{\pm}$, K$^{\pm}$, K$_{S}^{0}$ and $\bar{p}$). In 5-30\% central collisions
a sizable difference is present between the $v_1(y)$ slope of protons and antiprotons,
with the former being consistent with zero within errors. The $v_1$ excitation
function is presented. Comparisons to model calculations (RQMD,
UrQMD, AMPT, QGSM with parton recombination, and a hydrodynamics
model with a tilted source) are made. For those models which have calculations of $v_1$ for both pions and protons, none of them can
describe $v_1(y)$ for pions and protons simultaneously. The hydrodynamics model with a tilted source as currently implemented cannot explain
the centrality dependence of the difference between the $v_1(y)$
slopes of protons and antiprotons.
\end{abstract}

\maketitle

The BNL Relativistic Heavy Ion Collider
(RHIC) was built to study a new form of matter known as the Quark
Gluon Plasma (QGP)~\cite{whitePapers}, which existed in the universe shortly after the Big-Bang.
At RHIC, two nuclei are  collided at near light-speed,
and the collision produces thousands of particles due to the
significant energy deposited. The
collective motion of the produced particles can be
characterized~\cite{methodPaper} by Fourier coefficients,
\begin{linenomath}
\be v_{n} = \la  \cos n (\phi - \psi) \ra \label{vndef} \ee
\end{linenomath}
where $n$ denotes the harmonic, $\phi$ and $\psi$ denote the azimuthal angle of an outgoing particle
and reaction plane, respectively. The reaction plane
is defined by the collision axis and the line connecting the centers of two nuclei. 
Thus far, five of these coefficients have been measured and found to be non-zero
at RHIC~\cite{VoloshinPoskanzerSnellings}. They are directed flow $v_1$,
elliptic flow $v_2$, triangular flow $v_3$, the 4$^{th}$ order harmonic
flow $v_4$ and the 6$^{th}$ order harmonic flow $v_6$. This paper
will focus on the directed flow, the first Fourier coefficient.

Directed flow describes the sideward motion of produced particles in
ultra-relativistic nuclear collisions. It is believed to be generated during the
nuclear passage time before the thermalization happens, thus it carries early
information from the collision~\cite{Schnedermann,Kahana,Barrette,NA44}. The shape of
directed flow at midrapidity may be modified by the collective expansion and reveal a signature of a possible
phase transition from normal nuclear matter to
a QGP~\cite{antiFlow,thirdFlow,Stocker}. It is argued that
directed flow, as an odd function of rapidity $(y)$, may exhibit a small
slope (flatness) at midrapidity due to a strong expansion of the
fireball being tilted away from the collision axis. Such tilted
expansion gives rise to anti-flow~\cite{antiFlow} or a 3$^{rd}$
flow~\cite{thirdFlow} component (not the third flow harmonic). The anti-flow (or the 3$^{rd}$ flow component) is perpendicular to
the source surface, and is in the opposite direction to the
bouncing-off motion of nucleons. If the tilted expansion is strong
enough, it can even overcome the bouncing-off motion and results in
a negative $v_{1}(y)$ slope at midrapidity, potentially producing a
wiggle-like structure in $v_{1}(y)$. Note that although
calculations~\cite{antiFlow,thirdFlow} for both anti-flow and
3$^{rd}$ flow component are made for collisions at SPS energies
where the first order phase transition to a QGP is believed to be
the most relevant~\cite{Stocker}, the direct cause of the negative
slope is the strong, tilted expansion, which is also important at
RHIC's top energies. Indeed hydrodynamic calculations~\cite{Bozek}
for Au + Au collisions at $\sqrtsNN = 200$ GeV with a tilted source
as the initial condition can give a similar negative $v_1(y)$ slope
as that found in data.  A wiggle structure is also seen in the
Relativistic Quantum Molecular Dynamics (RQMD)
model~\cite{wiggleRQMD}, and it is attributed to baryon stopping
together with a positive space-momentum correlation. In this
picture, no phase transition is needed, and pions and nucleons flow
in opposite directions. To distinguish between baryon stopping and
anti-flow, it is desirable to
measure the $v_1(y)$ for identified particles and compare the sign
of their slopes at midrapidity. In particular, the observation of a
centrality dependence of proton $v_1(y)$ may reveal the character of
a possible first order phase transition~\cite{Stocker}. It is
expected that in very peripheral collisions the bouncing-off motion dominates over the entire rapidity range, 
and protons at midrapidity flow in the
same direction as spectators. In mid-central collisions, if there is
a phase transition, the proton $v_1(y)$ slope at midrapidity may
change sign and become negative. Eventually the slope diminishes in
central collisions due to the symmetry of the collisions.

At low energies, the E895 collaboration has shown that K$^0_S$ has a
negative $v_1(y)$ slope around midapidity~\cite{e895KShort}, while
$\Lambda$ and protons have positive slopes~\cite{e895Lambda}. This
is explained by a repulsive kaon-nucleon potential and an attractive
$\Lambda$-nucleon potential. The NA49 collaboration~\cite{na49} has
measured $v_1$ for pions and protons, and a negative $v_1(y)$ slope
is observed by the standard event plane method. The three-particle
correlation method $v_1\{3\}$~\cite{v1Cumu}, which is believed to be
less sensitive to non-flow effects, gives a negative slope too, but
with a larger statistical error. The non-flow effects are
correlations among particles that are not related to the reaction
plane, including the quantum Hanbury Brown-Twiss
correlation~\cite{HBT}, resonance decays~\cite{decay} and so on. At
top RHIC energies, $v_1$ has been studied mostly for charged
particles by both the STAR and the PHOBOS
collaborations~\cite{starV1V4,phobosV1,star62GeV,starV1PRL}. It is
found that $v_1$ in the forward region follows the limiting
fragmentation hypothesis~\cite{Limit}, and $v_1$ as a function
of pseudorapidity ($\eta$) depends only on the
incident energy, but not on the size of the colliding system at a
given centrality. Such system size
independence of $v_1$ can be explained by the hydrodynamic
calculation with a tilted initial condition~\cite{Bozek}. The
systematic study of $v_1$ for identified particles at RHIC did not
begin until recently because it is more challenging for two reasons:
1) $v_1$ for some identified particles (for example, protons) is
much smaller than that of all charged particles, 2) more statistics are needed to
determine $v_1$ for identified particles other than pions.

54 million events from Au + Au collisions at $\sqrtsNN = 200$
GeV have been used in this study, all taken by a minimum-bias
trigger with the STAR detector during RHIC's seventh run in year
2007. The main trigger detector used is the Vertex Position Detector
(VPD)~\cite{VPD}. The centrality definition of an event was based on
the number of charged tracks in the Time Projection Chamber
(TPC)~\cite{TPC} with track quality cuts: $|\eta|<$ 0.5, a Distance
of Closest Approach (DCA) to the vertex less than 3 cm, and
10 or more fit points. In the analysis, events are required to have
the vertex \textit{z} within 30 cm from the center of the TPC,
and additional weight is assigned to each event in the analysis
accounting for the non-uniform VPD trigger efficiency in the vertex
\textit{z} direction for different centrality classes. The event
plane angle is determined from the sideward deflection of spectator
neutrons measured by STAR's Shower Maximum Detector inside the Zero
Degree Calorimeters (ZDC-SMDs). Such sideward
deflection of spectator neutrons is expected to happen in
the reaction plane rather than participant plane, since the ZDC-SMDs
 are located close to beam rapidity. Being 6 units in $\eta$ away from
midrapidity, ZDC-SMDs also allow a measurement of $v_{1}$ with minimal
contribution from non-flow correlations. The description of measuring
$v_1$ using the ZDC-SMDs event plane can
be found in~\cite{star62GeV}. Particle Identification (PID) of
charged particles is achieved by measuring ionization energy loss
(\textit{dE/dx}) inside STAR's TPC, together with the measurement of the
momentum ($p$) via TPC tracking. Track quality cuts are the same as
used in~\cite{starFlowPRC}. In addition, the transverse momentum
$p_T$ for protons is required to be larger than 400 MeV$/c$, and DCA
is required to be less than 1 cm in order to
avoid including background protons which are from knockout/nuclear
interactions of pions with inner detector material. The same cuts
are applied to antiprotons as well to ensure a fair comparison with
protons. The high-end of the $p_T$ cut is 1 GeV$/c$ where protons
and pions have the same energy loss in the TPC and thus become
indistinguishable. For pions and kaons, $p_T$ range is
0.15 - 0.75 GeV$/c$ and 0.2 - 0.6 GeV$/c$, respectively. K$_S^0
(\rightarrow \pi^+ \pi^-)$ are topologically reconstructed by their
charged daughter tracks inside the TPC~\cite{KsReconstruction}.

Results presented in the following figures contain only statistical
errors. Results for pions, protons and antiprotons are not corrected
for the feeddown from weak decay particles. The major systematic
error in determining the slope of $v_1(y)$ for identified particles
is from the particle misidentification, which was evaluated by
varying the \textit{dE/dx} cut.
Another systematic error comes from the non-uniform $p_T$ acceptance,
as $v_1(y)$ is obtained by integrating $v_1$ over the $p_T$ acceptance
which itself depends on the rapidity. This effect is non-negligible
for protons and antiprotons at large rapidity. It is estimated by taking
the difference between slopes fitted with points integrated with $p_T$
acceptance at midrapidity and at large rapidity. In addition,
some of the observed protons have originated from interactions
between the produced particles and the detector material, and such effect
has also been taken into consideration. The total systematic
uncertainty is obtained by adding uncertainties mentioned above in quadrature. There are also
common systematic errors that should be applied to all particles:
the uncertainty due to the first order event plane determination, which was estimated to be
$\sim 10$\% (relative error) ~\cite{star62GeV}, and the uncertainty due to centrality selection, which was
estimated to be $\sim 4$\% (relative error) by comparing our charged
$v_{1}(\eta)$ slope to that from the RHIC run in 2004. Other
systematic errors have been evaluated to be negligible.

\begin{figure}
\resizebox{0.55\textwidth}{!}{%
  \includegraphics{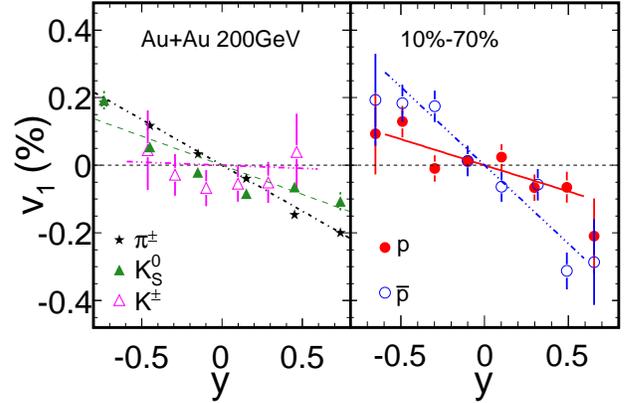}
}
\vspace{0.01cm}       
\caption{$v_1$ for $\pi^{\pm}$, K$^{\pm}$, K$_S^0$~(left panel), $p$
and $\bar{p}$~(right panel) as a function of rapidity for 10-70\% Au
+ Au collisions at $\sqrtsNN = 200$ GeV. The lines present the
linear fit to the $\pi^{\pm}$, K$^{\pm}$, K$_S^0$, $p$ and
$\bar{p}$'s $v_1(y)$ respectively. Data points around $y = 0.29$ are
slightly shifted horizontally to avoid overlapping.}
\label{fig:v1PID}       
\end{figure}

\begin{figure}
\resizebox{0.55\textwidth}{!}{%
  \includegraphics{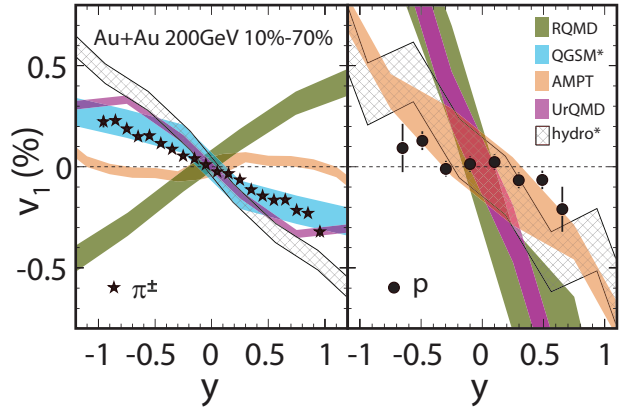} }
\vspace{0.01cm}       
\caption{Model calculations of pion (left panel) and proton (right
panel) $v_1(y)$ for 10-70\% Au + Au collisions at $\sqrtsNN = 200$
GeV. QGSM* model presents the basic Quark-Gluon String model with
parton recombination~\cite{QGSM}. Hydro* model presents  the
hydrodynamic expansion from a tilted source~\cite{Bozek}.}
\label{fig:v1PionProton}       
\end{figure}

In Fig.~\ref{fig:v1PID}, $v_1(y)$ of $\pi^{\pm}$, K$^{\pm}$,
K$_S^0$, $p$, and $\bar{p}$ are presented for centrality
10-70\%. Following convention, the sign of spectator $v_1$ in the
forward region is chosen to be positive, to which the measured sign
of $v_1$ for particles of interest is only relative. Fitting with a
linear function, the slopes are $-0.15 \pm 0.05 \text{(stat)} \pm
0.08\text{(sys)(\%)}$ for the protons, $-0.46 \pm 0.06 \text{(stat)}
\pm 0.04 \text{(sys)(\%)}$ for the antiprotons, $-0.27 \pm
0.01\text{(stat)} \pm 0.01\text{(sys)(\%)}$ for the pions, $-0.02
\pm 0.11 \text{(stat)} \pm 0.04 \text{(sys)(\%)}$ for the kaons and
$-0.17 \pm 0.02 \text{(stat)} \pm 0.04 \text{(sys)(\%)} $ for the
K$_S^0$. The relative $10\%$ common systematic error for all
particles is not listed here.
 The $v_1(y)$ slope for the produced particle types
($\pi^{\pm}$, K$^{\pm}$, K$_S^0$ and $\bar{p}$) are mostly found to
be negative at mid-rapidity, which is consistent with the anti-flow
picture. In particular, kaons are less sensitive to shadowing
effects due to the small kaon-nucleon cross section, yet it shows a
negative slope. This is again consistent with the anti-flow picture.
Interestingly, $v_{1}(y)$ for protons exhibits a clearly flatter
shape than that for antiprotons. While mass may contribute to the
difference in slope between pions and protons/antiprotons, it cannot
explain the difference in slope observed for antiprotons and
protons. Indeed, the observed $v_1$ for protons is a convolution of
directed flow of produced protons with that of transported protons
(from the original projectile and target nuclei), so the flatness of
inclusive proton $v_1(y)$ around midrapidity could be explained by
the negative flow of produced protons being compensated by the
positive flow of protons transported from spectator rapidity, as a
feature expected in the anti-flow picture.

In Fig.~\ref{fig:v1PionProton}, pion and proton $v_1(y)$
are plotted together with five model
calculations, namely, RQMD~\cite{wiggleRQMD}, UrQMD~\cite{UrQMD},
AMPT~\cite{AMPT}, QGSM with parton recombination~\cite{QGSM},
and slopes from an ideal hydrodynamic calculation with a tilted
source~\cite{Bozek}. The model calculations are performed in the
same $p_{T}$ acceptance and centrality as the data. The RQMD and
AMPT model calculations predict the wrong sign and wrong magnitude of pion $v_1(y)$, respectively,
while the RQMD and the UrQMD predict the
wrong magnitude of proton $v_1(y)$. For models other than QGSM which has the calculation only for pions, none of them can describe $v_1(y)$ for pions and
protons simultaneously.

\begin{figure}
\resizebox{0.50\textwidth}{!}{%
  \includegraphics{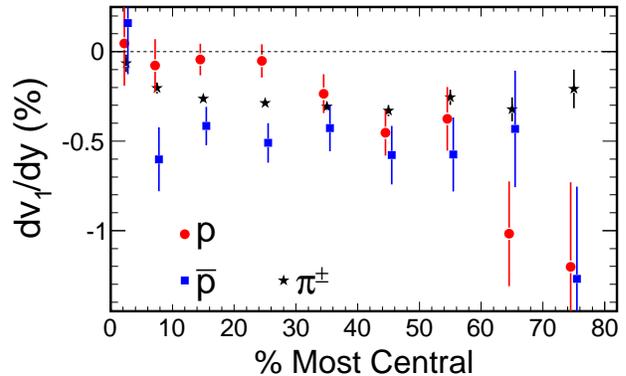}
}
\vspace{0.01cm}       
\caption{Charged pions~(solid stars), protons~(solid circles) and
antiprotons~(solid squares) $v_1(y)$ slope~($dv_1/dy$) at
midrapidity as a function of centrality for Au + Au collisions at
$\sqrtsNN = 200$ GeV. }
\label{fig:v1Slope_cent}       
\end{figure}

In Fig.~\ref{fig:v1Slope_cent}, the slope of $v_1(y)$ at midrapidity is presented as a function of
centrality for protons, antiprotons, and charged pions. In general, the magnitude of the $v_{1}(y)$
slope converge to zero as expected for most central collisions. Proton and antiproton $v_1(y)$ slope are more or less consistent in 30-80\% centrality range but, diverge in 5-30\% centrality. In addition, two
observations are noteworthy: i)~the hydrodynamic model with tilted
source (which is a characteristic of anti-flow) as currently implemented does not predict the
difference in $v_{1}(y)$ between particle
species~\cite{PiotrPrivate}. ii)~If the difference between $v_1$ of
protons and antiprotons is caused by anti-flow alone, then such
difference is expected to be accompanied by strongly negative $v_1$
slopes. In data, the large difference between proton and antiproton
$v_1$ slopes is seen in the 5-30\% centrality range,
while strongly negative $v_1$ slopes are found for protons,
antiprotons and charged pions in a different centrality
range (30-80\%). Both observations suggest that additional mechanisms than that assumed in~\cite{Bozek,PiotrPrivate} are needed to explain the
centrality dependence of the difference between the $v_1(y)$ slopes
of protons and antiprotons.

\begin{figure}
\resizebox{0.50\textwidth}{!}{%
  \includegraphics{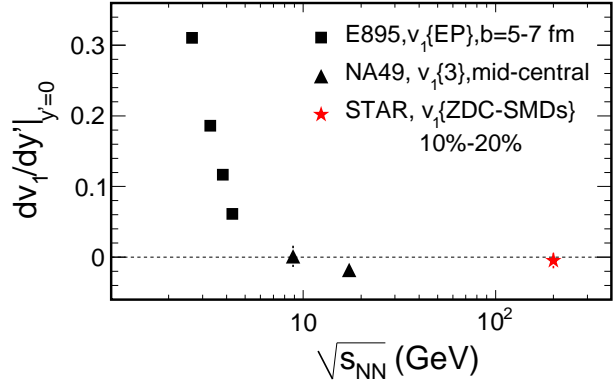}
}
\vspace{0.01cm}       
\caption{Proton $v_1(y^{'})$ slope ($dv_1/dy^{'}$) at midrapidity
as a function of center of mass collision energy, where $y^{'}=y/y_{beam}$. }
\label{fig:v1ExcitationFcn}       
\end{figure}

The excitation function of proton $v_1(y')$ slope $F$
($=dv_1/dy'$ at midrapidity) is presented in Fig~\ref{fig:v1ExcitationFcn}. Values for $F$ are extracted via a
polynomial fit of the form $Fy'+Cy'^{3}$, where $y^{'}=y/y_{beam}$ for which spectators are normalized at $\pm$1.
The proton $v_1(y')$ slope decreases rapidly with increasing energy,
reaching zero around $\sqrtsNN = 9$ GeV. Its sign changes to
negative as shown by the data point at $\sqrtsNN = 17$ GeV, measured
by the NA49 experiment~\cite{na49}. A similar trend has been
observed at low energies with a slightly different quantity $d\langle p_{x}\rangle/dy'$~\cite{E877Slope,E895PRL}. 
The energy dependence of $v_1(y')$
slope for protons is driven by two factors, i)~the increase in the
number of produced protons over transported protons with increasing
energy, and ii)~the $v_1$ of both produced and transported protons
at different energies. The negative $v_1(y')$ slope for protons
around midrapidity at SPS energies cannot be explained by transport
model calculations like UrQMD~\cite{Zhu2006} and AMPT~\cite{AMPT},
but is predicted by hydro calculations~\cite{antiFlow, thirdFlow}.
The present data indicate that the proton $v_1$ slope remains
close to zero at $\sqrtsNN = 200$ GeV as observed at
$\sqrtsNN = 9$ GeV and $\sqrtsNN = 17$ GeV heavy ion collisions.
Our measurement offers a unique check
of the validity of a tilted expansion at RHIC top energy.

In summary, STAR's measurements of directed flow of pions,
kaons, protons, and antiprotons for Au + Au collisions at
$\sqrtsNN = 200$ GeV are presented. In the range of 10-70\%
central collisions, $v_1(y)$ slopes of pions, kaons~(K$_S^0$), and
antiprotons are found to be mostly negative at mid-rapidity.
In 5-30\% central collisions
a sizable difference is present between the $v_1(y)$ slope of protons and antiprotons,
with the former being consistent with zero within errors. Comparison to models (RQMD, UrQMD, AMPT,
QGSM with parton recombination, and hydrodynamics with a
tilted source) is made. Putting aside the QGSM model which has the calculation only for pions, none of the other models explored can
describe $v_1(y)$ for pions and protons simultaneously.
Additional mechanisms than that assumed in the hydrodynamic model with a tilted source~\cite{Bozek,PiotrPrivate} are needed to explain
the centrality dependence of the difference between the $v_1(y)$
slopes of protons and antiprotons. Our measurement indicates
that the proton's $v_1(y)$ slope remains close to zero for Au + Au collisions at $\sqrtsNN = 200$ GeV. These new
measurements on the particle species and centrality dependence
of $v_1(y)$ provides a check for the validity of a tilted expansion 
at RHIC top energy.


\begin{acknowledgments}
We thank the RHIC Operations Group and RCF at BNL, the NERSC Center at LBNL and the Open Science Grid consortium for providing resources and support. This work was supported in part by the Offices of NP and HEP within the U.S. DOE Office of Science, the U.S. NSF, the Sloan Foundation, the DFG cluster of excellence `Origin and Structure of the Universe'of Germany, CNRS/IN2P3, FAPESP CNPq of Brazil, Ministry of Ed. and Sci. of the Russian Federation, NNSFC, CAS, MoST, and MoE of China, GA and MSMT of the Czech Republic, FOM and NWO of the Netherlands, DAE, DST, and CSIR of India, Polish Ministry of Sci. and Higher Ed., Korea Research Foundation, Ministry of Sci., Ed. and Sports of the Rep. Of Croatia, and RosAtom of Russia.
\end{acknowledgments}


\end{document}